\documentclass[12pt]{spie}  % 12pt font required by SPIE;
\usepackage{amsmath,amsfonts,amssymb}
\usepackage{graphicx}
\usepackage{setspace}
\usepackage{tocloft}
\newcommand{\tu}{\textunderscore}

\title{Survey strategy optimization for the Atacama Cosmology Telescope}

\author[a]{F.~De~Bernardis}
\author[a]{J.~R.~Stevens}
\author[b,c]{M.~Hasselfield}
\author[d]{D.~Alonso}
\author[e]{J.~R.~Bond}
\author[d]{E. Calabrese}
\author[f]{S.~K.~Choi}
\author[f]{K.~T.~Crowley}
\author[g]{M.~Devlin}
\author[d]{J.~Dunkley}
\author[a]{P.~A.~Gallardo}
\author[a]{S.~W.~Henderson}
\author[h]{M.~Hilton}
\author[i]{R.~Hlozek}
\author[f]{S.~P.~Ho}
\author[j]{K.~Huffenberger}
\author[a]{B.~J.~Koopman}
\author[k]{A.~Kosowsky}
\author[l]{T.~Louis}
\author[m]{M.~S.~Madhavacheril}
\author[n]{J.~McMahon}
\author[d]{S.~Naess}
\author[g]{F.~Nati}
\author[i]{L.~Newburgh}
\author[a]{M.~D.~Niemack}
\author[f]{L.~A.~Page}
\author[f]{M.~Salatino}
\author[o]{A.~Schillaci}
\author[g]{B.~L.~Schmitt}
\author[m]{N.~Sehgal}
\author[h]{J.~L.~Sievers}
\author[f]{S.~M.~Simon}
\author[p]{D.~N.~Spergel}
\author[f]{S.~T.~Staggs}
\author[e]{A.~van~Engelen}
\author[a]{E.~M.~Vavagiakis}
\author[q]{E.~J.~Wollack}

\affil[a]{Cornell University, Physics Department, Ithaca, NY 14850, USA}
\affil[b]{Department of Astronomy and Astrophysics, The Pennsylvania State University, University Park, PA, 16802, USA}
\affil[c]{Institute for Gravitation and the Cosmos, The Pennsylvania State University, University Park, PA 16802, USA}
\affil[d]{University of Oxford, Denys Wilkinson Building, Keble Road, Oxford, OX1 3RH, UK}
\affil[e]{CITA, University of Toronto, 60 St. George St., Toronto, ON M5S 3H8, Canada}
\affil[f]{Department of Physics, Princeton University, Princeton, New Jersey, 08544 USA}
\affil[g]{Department of Physics and Astronomy, University of Pennsylvania, Philadelphia, PA 19104, USA}
\affil[h]{University of KwaZulu-Natal, Westville Campus, Durban 4041, ZA}
\affil[i]{Dunlap Institute for Astronomy \& Astrophysics, University of Toronto, Toronto, ON M5S 3H4, Canada}
\affil[j]{Florida State University, Tallahassee, FL 32306, USA}
\affil[k]{Department of Physics and Astronomy, University of Pittsburgh, Pittsburgh, PA 15260 USA}
\affil[l]{UPMC Univ Paris 06, UMR7095, Institut d’Astrophysique de Paris, F-75014, Paris, France}
\affil[m]{Stony Brook University, Stony Brook, NY 11794}
\affil[n]{Department of Physics, University of Michigan Ann Arbor, MI 48109, USA}
\affil[o]{Instituto de Astrofisica and Centro de Astro-Ingenieria, Facultad de Fisica, Pontiﬁcia Universidad Cattolica de Chile, Santiago, Chile}
\affil[p]{Dept. of Astrophysical Sciences, Peyton Hall, Princeton University, Princeton, NJ USA 08544}
\affil[q]{NASA Goddard Space Flight Center, 8800 Greenbelt Road, Greenbelt, Maryland 20771, USA}

\cftpagenumbersoff{figure}
\cftpagenumbersoff{table} 
\begin{document} 
\maketitle

\begin{abstract}
In recent years there have been significant improvements in the sensitivity and the angular resolution of
the instruments dedicated to the observation of the Cosmic Microwave Background (CMB). ACTPol is
the first polarization receiver for the Atacama Cosmology Telescope (ACT) and is observing the CMB
sky with arcmin resolution over $\sim$2000 sq. deg. Its upgrade, Advanced ACTPol (AdvACT), will observe the CMB
in five frequency bands and over a larger area of the sky. We describe the optimization and
implementation of the ACTPol and AdvACT surveys. The selection of the observed fields is
driven mainly by the science goals, that is, small angular scale CMB measurements, B-mode measurements and
cross-correlation studies. For the ACTPol survey we have observed patches of the southern galactic sky
with low galactic foreground emissions which were also chosen to maximize the overlap with several galaxy surveys to
allow unique cross-correlation studies. A wider field in the northern galactic cap ensured
significant additional overlap with the BOSS spectroscopic survey. The exact shapes and footprints of
the fields were optimized to achieve uniform coverage and to obtain cross-linked maps by observing the
fields with different scan directions. We have maximized the efficiency of the survey by implementing a
close to 24 hour observing strategy, switching between daytime and nighttime observing plans and
minimizing the telescope idle time. We describe the challenges represented by the survey optimization
for the significantly wider area observed by AdvACT, which will observe roughly half of the low-foreground sky. 
The survey strategies described here may prove useful for planning future ground-based CMB surveys, such as the Simons Observatory and  CMB Stage IV surveys. 
%and the software package used to optimize and implement the ACT surveys.

\end{abstract}

\section{Introduction}
The Cosmic Microwave Background (CMB) remains one of the most valuable sources of cosmological information. Temperature anisotropy measurements have reached the cosmic variance limit at angular scales $\gtrsim 0.1^{\circ}$ with the Planck satellite\cite{Ade:2015xua}. Polarization measurements can provide important additional information able to break degeneracies between cosmological parameters and constrain extensions of the $\Lambda$CDM model, such as the tensor-to-scalar ratio $r$ and the neutrino mass sum $\sum m_{\nu}$. Several ground based experiments, such as the Atacama Cosmology Telescope Polarimeter (ACTPol) \cite{2010SPIE.7741E..1SN}, POLARBEAR \cite{2010arXiv1011.0763T}, SPTpol \cite{2012SPIE.8452E..1EA}, BICEP2 \cite{Ade:2014gua}, Keck-array \cite{2012JLTP..167..827S} and CLASS \cite{2014SPIE.9153E..1IE},  are measuring the E-mode and B-mode polarization signals \cite{Hanson:2013hsb, Ade:2014xna, 2014ApJ...794..171T, 2014JCAP...10..007N}.

In this paper we focus on the survey strategy implemented by ACTPol between 2013 and 2015 and the upgraded instrument AdvACT \cite{Henderson:2015nzj}, which started its first observation campaign in 2016. AdvACT will observe approximately half of the sky in five frequency bands, from 28 GHz to 230 GHz, with the high angular resolution of 1.4 arcmin at 150 GHz already achieved by ACTPol. The expected map noise in temperature and polarization will be significantly reduced with respect to the ACTPol survey thanks to the nearly doubled number of detectors. AdvACT plans to use half-wave plates (HWP) that modulate the polarized signal at several Hz to improve polarization measurements at the largest angular scales.

One of the unique advantages of the ACTPol and AdvACT surveys is the large overlap with optical surveys like BOSS\cite{2013AJ....145...10D}, HSC\cite{HSC2006}, DES\cite{DES}, DESI\cite{Levi:2013gra} and LSST\cite{Ivezic:2008fe}. This overlap allows for powerful cross-correlation studies and new probes of dark energy and the neutrino mass sum. %CMB temperatures and polarizations measurements up to small angular scales will enable improved measurements of the CMB lensing and cross-correlations studies will be able to detect secondary anisotropies like the thermal and kinematic Sunyaev-Zel'dovich effects, allowing for new powerful probes for the dark energy parameters and the neutrino mass.

The observation plans for ACTPol and AdvACT are designed to maximize the scientific potential of the surveys. The choice of the observed fields and of the relative priority for different fields (i.e. which field is observed when more than one field is visible at the same time) takes into account several constraints and scientific objectives that often compete with each other. Observing regions that maximize overlap with optical surveys is an important goal for cross-correlation studies. Galactic dust emission is a significant limitation especially for measurements of large scale B-mode polarization, which thus favor observations of low foreground regions. Scanning the fields while rising and setting at different elevations (cross-linking) reduces systematic effects in the map-making process\cite{Sutton:2008zh}. For the daytime measurements, sun avoidance introduces an additional constraint that must be accounted for, or significant data loss can result from data acquired with the sun in the sidelobes. Moon avoidance is less concerning. We have not yet accounted for it in the ACTPol strategy nor in the 2016 AdvACT strategy but we plan to include it in future observing plans. We take these constraints into account and maximize the efficiency of the observing plan by minimizing the idle time of the telescope and switching between different observing plans for daytime and nighttime. 

In section \ref{sec_actpol} we describe the three seasons of observations with ACTPol from 2013 to 2015 while in section \ref{sec_advactpol} we focus on the survey strategy for AdvACT discussing the challenges represented by the much wider area covered by the AdvACT survey.

\section{ACTPol survey}\label{sec_actpol}
The first season of observations with the ACTPol receiver (season 1) was conducted in 2013. ACTPol observed four deep patches centered near $0^{\circ}$ declination (Dec) and at right ascensions (RA) $150^{\circ}$ (\textit{deep1}), $175^{\circ}$ (\textit{deep2}), $355^{\circ}$ (\textit{deep5}) and $35^{\circ}$ (\textit{deep6}) with areas of $73$, $70$, $70$ and $63$ sq. deg. An additional field located on the galactic plane at (RA, Dec) = ($287^{\circ}$,$0^{\circ}$) was observed as a lower priority daytime field for the last part of the season.  These fields were selected based on their overlap with other surveys. The exact location of these fields in the sky were optimized so that only one patch would be visible at any given time. Some details about this strategy were presented with temperature and polarization power spectra from the first season of observations in Naess et al. 2014\cite{2014JCAP...10..007N}. Here we focus on the survey strategies for the 2014 and 2015 observations with ACTPol. 
\begin{figure}[!h]
\begin{center}
\begin{tabular}{c}
\includegraphics[height=9cm]{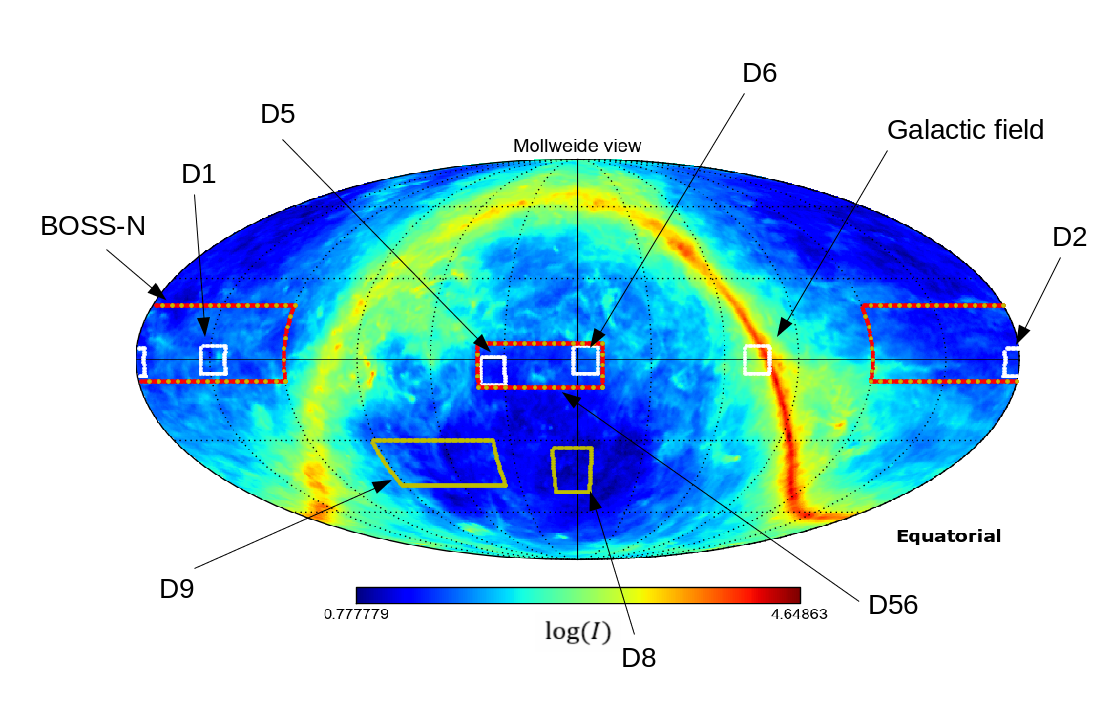}
\end{tabular}
\end{center}
\caption{Fields observed in the three seasons of observations with ACTPol, in equatiorial coordinates with RA increasing from the center (RA = $0^{\circ}$) to the left. The white fields were observed in season 1, red in season 2 and yellow in season 3. The background shows the thermal dust component emission ($log_{10}$ of the intensity at $545$ GHz in units of  $\mu K_{RJ}$) as measured by the Planck satellite \cite{planckmaps}.} \label{ACTPol_seasons}
\end{figure}

Figure \ref{ACTPol_seasons} shows the fields observed in the three ACTPol seasons from 2013 to 2015. In the 2014 season (season 2) ACTPol observed two regions centered at (RA, Dec) $= (16^{\circ},-2^{\circ})$ and (RA,Dec) $= (180^{\circ},8^{\circ})$, called respectively \textit{deep56} and \textit{BOSS-N}. The field \textit{deep56} covers $\sim 700$ sq. deg. and encloses the \textit{deep5} and \textit{deep6} fields observed during season 1. This field is the deepest patch observed during season 2 and it was selected to obtain overlap with several galaxy surveys, including the BOSS southern galactic field, HSC and DES while at the same time increasing the sensitivity in the \textit{deep5} and \textit{deep6} fields already observed during season 1. The second field (\textit{BOSS-N}) covers about $\sim 2000$ sq. deg. and it was selected to achieve overlap with the BOSS survey northern galactic field while also providing overlap with KIDs \cite{2015A&A...582A..62D} and HSC. 

\begin{figure}[!h]
\begin{center}\
 \begin{tabular}{c}
\includegraphics[height=9cm]{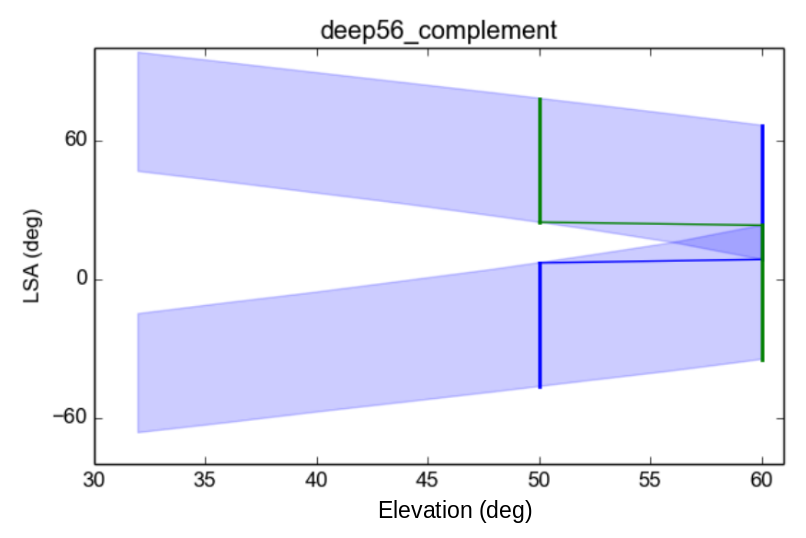}
\end{tabular}
\end{center}
\caption{Example of optimization of the survey strategy for deep56.  Shaded region shows combinations of local sidereal angle and telescope elevation during which deep56 is visible, where the lower and upper bands represent the rising and setting of the field, respectively.  The blue line shows a trajectory through the space that minimizes the temporal (LSA) gap between a rising and a setting observation of the field.  The green line shows a second such trajectory.  Such diagrams are used to optimize between field size and elevation of observation.}\label{cover}
\end{figure}

The ACTPol survey strategy consists of horizontal scans at different elevations. Each field is observed by scanning back-and-forth in azimuth. The total elevation range of ACT is $30^{\circ}$ to $60^{\circ}$. Observations at higher elevations are generally preferred to reduce the amount of atmosphere observed by the telescope and to avoid emissions from the ground. We observed \textit{deep56} at elevations of $50^{\circ}$ and $60^{\circ}$ both while rising and setting for each elevation. Figure \ref{cover} shows how the survey strategy for \textit{deep56} is optimized, by looking for a trajectory through the local sidereal angle (LSA)-elevation space that minimizes the temporal gap between a rising and a setting observation of the field. A similar optimization is implemented for the other fields. This configuration provides cross-linked maps of the field. Table \ref{seasons_params} summarizes the observing parameters for \textit{deep56} and \textit{BOSS-N} for seasons 2 and 3 with their respective elevations, central azimuths and azimuth ranges. The summary of the areas observed in the 2014 and 2015 seasons of ACTPol is presented in Table \ref{area_summary}. The \textit{BOSS-N} field is the widest field observed by ACTPol and it extends to the highest declinations. Given these characteristics achieving good cross-linking requires using lower elevations. Both rising and setting observations of \textit{BOSS-N} were made at $33^{\circ}$ and $38^{\circ}$ elevation. 

\begin{table}
\caption{Summary of the ACTPol surveys parameters for seasons 2 and 3 for the main fields observed during those seasons. We report elevations, east and west azimuth of observation (azE, azW) and azimuth scan amplitudes (throw) in degrees for BOSS-N, deep56, deep8 and deep9. The peak-to-peak azimuth range is twice the throw. Note that deep56 and BOSS-N were split into two strips for part of 2014 to reduce the azimuth throw.}\label{seasons_params}
\begin{center}
\begin{tabular}{ l | c | c | c | c | c  }
Year & Field  &  elevation ($^{\circ}$)    & azE ($^{\circ}$)   & azW($^{\circ}$)    & throw ($^{\circ}$)\\
\hline
2014-08-22 & BOSS-N-A & 33   & 54  & 305  & 10         \\
     & BOSS-N-B & 38   & 68  & 292  & 9         \\
     & deep56-A & 50   & 57  & 302  & 7         \\
     & deep56-A & 60   & 37  & 322  & 13         \\
     & deep56-B & 50   & 70  & 289  &  6  \\
     & deep56-B & 60   & 59  & 301  &   9  \\
\hline
2014-10-08 & BOSS-N-A & 33  & 54  & 305  & 10  \\
           & BOSS-N-B & 38  & 68  & 292  & 9  \\
           & deep56   & 50  & 63  & 297  & 13  \\
           & deep56   & 60  & 46  & 314   & 22  \\
\hline
2015 & BOSS-N &  38     &  56     & 303     & 20       \\
     & BOSS-N &  35     &  59     &  300    & 19   \\
     & deep56 &  50     &  62     & 297     & 13        \\
     & deep56 &  60     &  45     & 314     & 21      \\
     & deep8  &   35     &  126     & -     & 10        \\
     & deep8  &   49     &  128     & -     & 14      \\
     & deep9  &   35     &  237     & -     & 11      \\
     & deep9  &   60     &  229     & -     & 20      \\

\hline
\end{tabular}
\end{center}
\end{table}

For the first part of season 2 we split \textit{deep56} into two strips with parallel declination (deep56-A and deep56-B), overlapping $1^{\circ}$ in elevation and observed them in separated scans. One advantage of this approach is that since the azimuth scan amplitude was reduced the scans across the same field were faster. However, this declination splitting introduced more uneven coverage at the edge between the two strips, corresponding to the telescope turning around. For the rest of season 2 we observed \textit{deep56} as a single field using wider azimuth scans. \textit{BOSS-N} was also split into two strips (BOSS-N-A, BOSS-N-B) with parallel declinations to reduce the azimuth throw overlapping for about 1$^{\circ}$ at Dec = $7^{\circ}$.

\begin{figure}[!ht]
\centering
%\begin{minipage}{.5\textwidth}
%  \centering
  \includegraphics[width=1\linewidth]{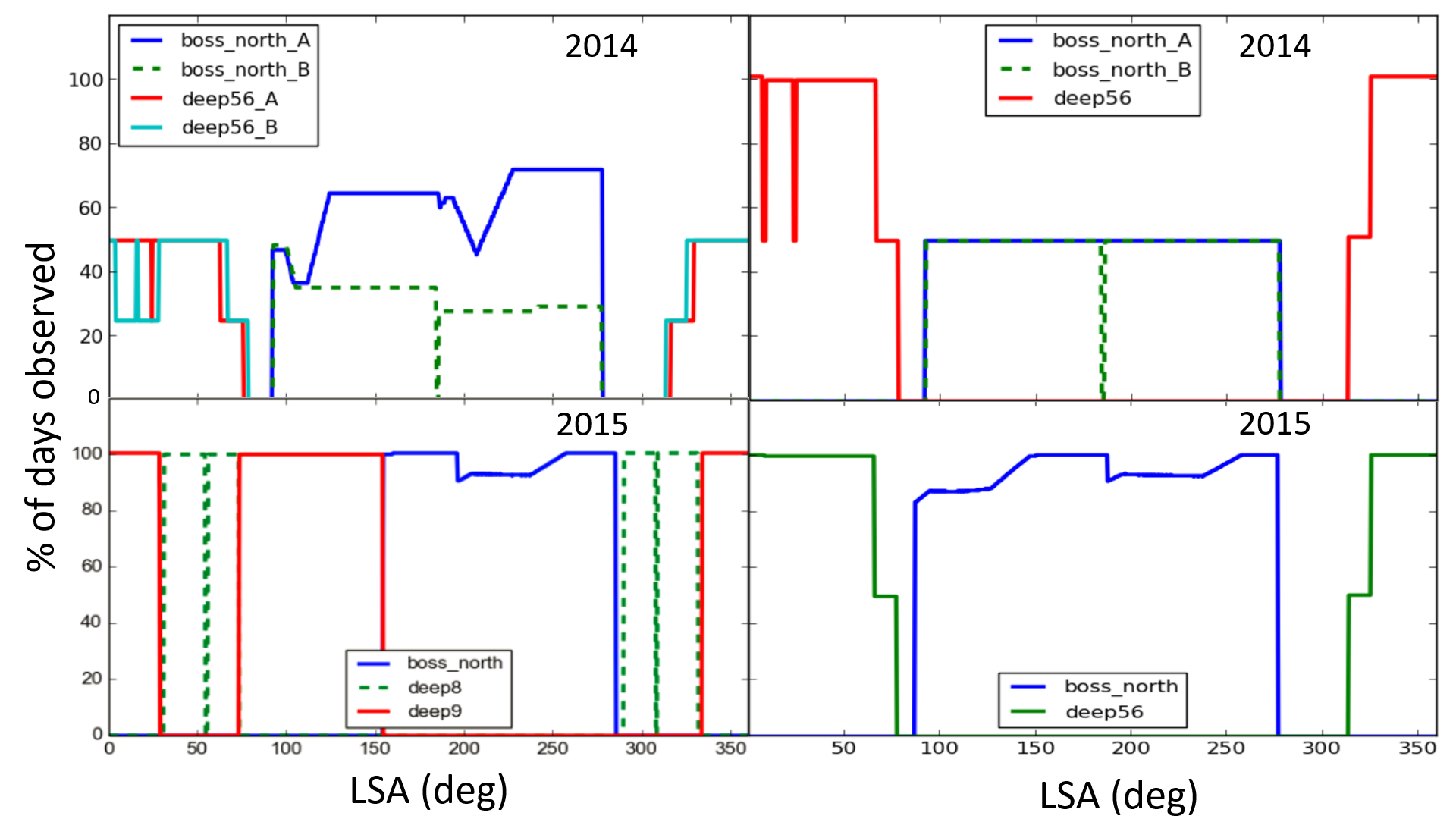}\hfill
  \caption{Fraction of days of the nominal plan allocated to observations of the various ACTPol fields at a given local sidereal time expressed as local sidereal angle (LSA). The fraction is calculated with respect to the total number of days of the nominal survey plan: a $100\%$ fraction indicates that the field was planned to be observed every day of the season. The fields are selected to complement and minimize the telescope idle time by filling most of the LSA range. Gaps in the LSA plot indicate that none of the fields are visible. A field might not be observed $100\%$ of the total available time to avoid the sun or if, for some time during the season, it is only visible for a time shorter than the minimum service time. The minimum service time is introduced to avoid uneven coverage at the edge of the fields. \textit{Top row:} fields observed during season 2. We started by splitting both deep56 and BOSS-N into two strips (left) between August and October. We continued observing deep56 as a single field (right) from October to December. \textit{Bottom row:} the two survey plans used for season 3. We started season 3 using the same strategy as season 2 (April-May). In May we merged these two plans into a single schedule using the first (left) as daytime plan and the second (right) as nighttime plan, observing from May to December. }\label{lsa_actpol}
\end{figure}

The choice of the elevations was constrained by the sizes and location of the fields and it was optimized so that the observing configurations (rising and setting at the chosen elevations) interlocked in time. Figure \ref{lsa_actpol} shows the fraction of days spent observing the various fields at a given local sidereal angle (LSA), emphasizes the complementarity of the chosen field and that in season 3 the gaps in the observing plan were minimized by merging a nighttime and a daytime strategy.

For the first few weeks of season 3, which began in April 2015, we used the same observing plan of season 2 but removed the declination splitting in \textit{BOSS-N}. For the second part (from May 11 to December 31) we introduced two additional fields, \textit{deep8} and \textit{deep9}, at (RA, Dec) = $(2.5^{\circ}, -42^{\circ})$ and $(65^{\circ},-39^{\circ})$ with areas $\sim 190$ sq. deg. and $\sim 700$ sq. deg.. These patches were selected to overlap with the lowest galactic foreground areas of the sky and the main goal will be to use them to reconstruct the low multipole part of the B-modes power spectra that is more affected by galactic foregrounds. The observing parameters for \textit{deep8} and \textit{deep9} are also shown in Table \ref{seasons_params}. To maximize the efficiency and achieve good coverage for \textit{deep56}, \textit{BOSS-N}, \textit{deep8} and \textit{deep9} we produced two independent survey plans, one prioritizing observations of \textit{deep8} and \textit{deep9} and one observing \textit{deep56} and \textit{BOSS-N} as higher priority fields. We then merged the two plans into a single schedule, using the first (\textit{deep8} and \textit{deep9}) as the daytime plan the second (\textit{deep56} and \textit{BOSS-N}) as the nighttime plan, where we define daytime as 12:00-22:00 local time. The reason for this splitting is the distortion of the beam shape observed during the day caused by thermal distortions of the telescope structure. The largest angular scales are less affected by the details of the beam shape, while the small angular scales, important for lensing reconstruction and cluster science, require good control of the beam. This is the motivation for prioritizing regions with good optical data for cross-correlations, \textit{deep56} and \textit{BOSS-N}, at night.

\begin{figure}[!h]
\begin{center}
\begin{tabular}{c}
\includegraphics[height=10cm]{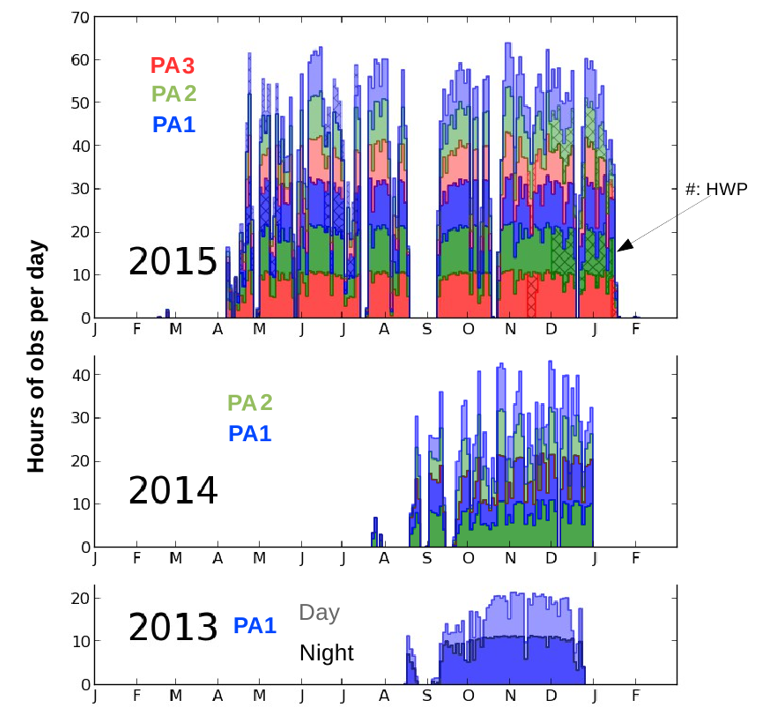}
\end{tabular}
\end{center}
\caption{Summary of the hours of observation per day for the three ACTPol seasons. The colors specify which polarimeter array (PA) hours are indicated. The darker shades of the color indicate night observations, while the light shades indicate day observations. Night is always below day. The hashed areas indicate observations with a half-wave plate. Each season incorporated a new polarization array (PA). The gap in September 2015 was due to a crystoat repair. The other minor gaps are due to weather unsuitable for observations.}\label{eff}
\end{figure}

Figure \ref{eff} shows a summary of the hours observed per day during each month of ACTPol seasons 1 to 3. We report the amount of observing time per detector array both for nighttime and daytime observations. 

\begin{figure}[!h]
\begin{center}
\begin{tabular}{c}
\includegraphics[height=8cm]{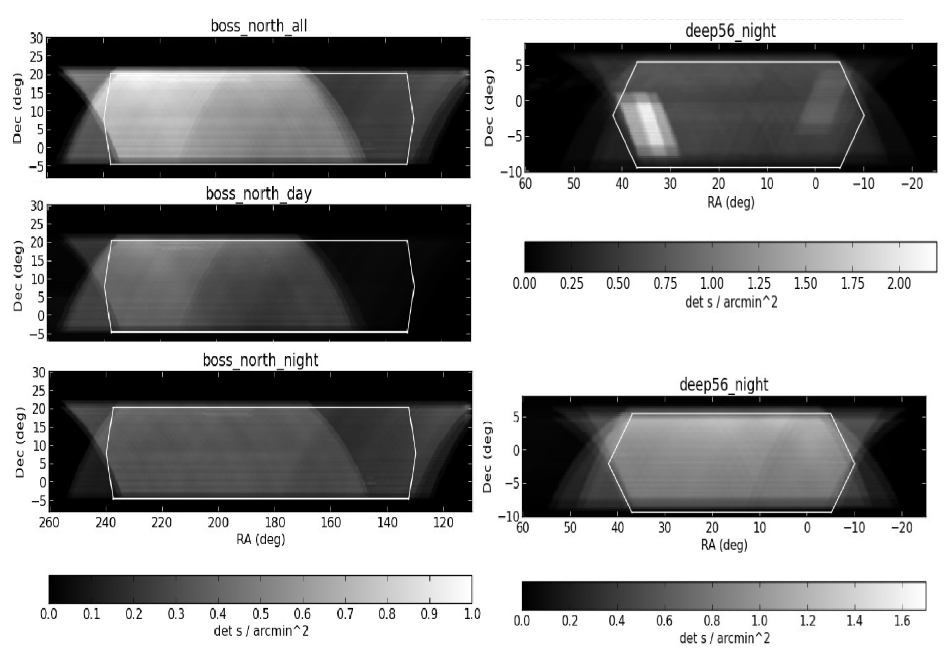}
\end{tabular}
\end{center}
\caption{\textit{Left}: Weight maps showing the coverage of the BOSS-N fields for season 3. The coverage are relative to night only (bottom), day only (middle) and total (top). \textit{Right}: Weight maps for deep56, observed mostly at night, during season 2 (top) where the deeper patches corresponding to deep5 and deep6 are clearly visible, compared with season 3 (bottom). The black and white scale shows the telescope time per area, i.e. assuming $N=1$ detectors. The white outlines define the areas used to guide the generation of the survey strategy.}\label{weights_actpol}
\end{figure}

\begin{figure}[!h]
\begin{center}
\begin{tabular}{c}
\includegraphics[height=5cm]{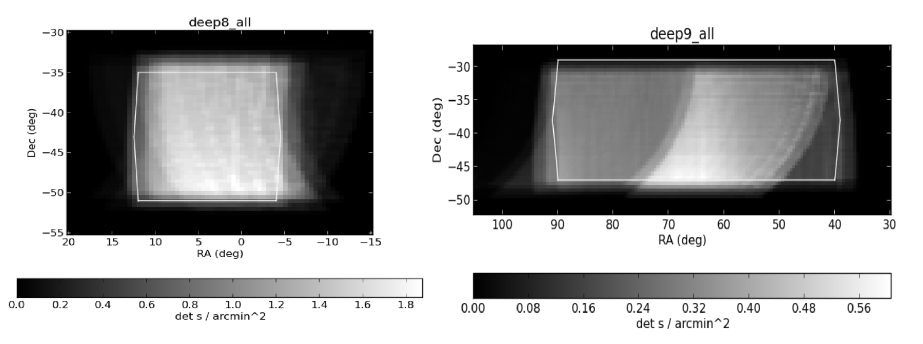}
\end{tabular}
\end{center}
\caption{Weight maps for the daytime fields deep8 and deep9 observed during season 2.}\label{deep8_deep9}
\end{figure}

In the last season ACTPol tested the polarization modulation, obtaining several hours of observations with the HWP installed in front of the optics tube. Figure \ref{weights_actpol} shows the coverage obtained in season 2 and 3 for \textit{deep56} and season 3 for \textit{BOSS-N} and Figure \ref{deep8_deep9} shows the season 3 coverage for \textit{deep8} and \textit{deep9}.  

\begin{table}
\caption{Summary of areas (sq. deg.) observed by ACTPol during season 2 and 3.}\label{area_summary}
\begin{center}
\begin{tabular}{ l | c | c | c | c }
           &  deep56   & BOSS-N   & deep8 & deep9  \\
\hline
Season 2 (2014)       &  700 sq. deg.      & 2000 sq. deg.   &    - & -      \\
\hline
Season 3 (2015)       &  700 sq. deg.    &  2000 sq. deg.  &  190 sq. deg.  & 700 sq. deg.      \\
 \hline
\end{tabular}
\end{center}

\end{table}

\section{AdvACT survey 2016}\label{sec_advactpol}
The AdvACT survey will significantly extend the area observed by ACTPol over the next five years, observing an area close to half the sky. The wider fields observed by AdvACT pose several challenges.

 Achieving uniform coverage for all the fields requires a finer optimization of the elevations and azimuth ranges and a more complicated succession of rising and setting observations for different fields on different days. The scans will be wider in azimuth but at the same time they need to be fast enough so that while the telescope returns to the original position the field has not moved substantially across the sky. We are continuing to study the optimal azimuth throw for calibration and to maintain uniform coverage of the fields.

\begin{figure}[!h]
\begin{center}\
 \begin{tabular}{c}
\includegraphics[height=9cm]{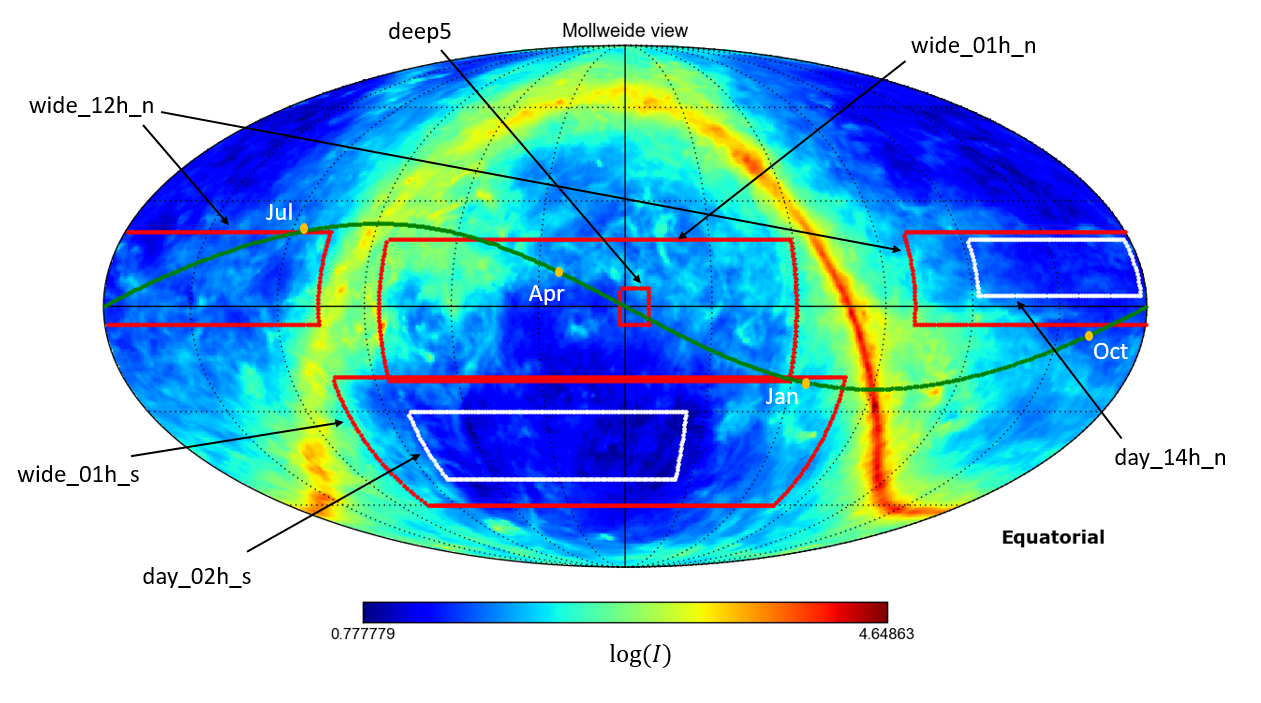}
\end{tabular}
\end{center}
\caption{Night time (red) and daytime (white) fields observed by the AdvACT survey. The green curve shows the Sun trajectory during the year. See Tables \ref{merging_sequence} and \ref{area_summary_advact} for more details about the fields and the observing plan. The background shows the thermal dust component emission ($log_{10}$ of the intensity at $545$ GHz in units of  $\mu K_{RJ}$) as measured by the Planck satellite \cite{planckmaps}.}\label{advact_fields}
\end{figure}

Similarly to ACTPol season 3 we split the observing plan into a nighttime and a daytime strategy. We dedicate the nighttime to observations of the widest field. Figure \ref{advact_fields} shows the nighttime and daytime fields. During nighttime we observe three wide fields, two centered at RA $\simeq 1^{\circ}$ and declinations Dec $0^{\circ}$ and $-40^{\circ}$  (\textit{wide\tu01h\tu n} and \textit{wide\tu01h\tu s}) with areas $\sim 5700$ sq. deg. and $\sim 5000$ sq. deg. and one at (RA, Dec) = (12$^{\circ}$, 8$^{\circ}$) (\textit{wide\tu12h\tu n}) covering $\sim 3400$ sq. deg.. The areas to be observed by AdvACT in 2016 are summarized in Table \ref{area_summary_advact}. These fields are observed at three different elevations, $40^{\circ}$, $45^{\circ}$ and $47.5^{\circ}$, both while rising and while setting. The fields \textit{wide\tu01h\tu n} and \textit{wide\tu01h\tu s}  cannot be observed at the same time, while \textit{wide\tu12h\tu n} can be observed in combination with one of the others. To achieve observations at multiple altitudes for all the wide fields and avoid potential systematic effects associated with any single altitude choice, we then produce twelve independent observing strategies, four for each elevation and alternating rising and setting every night according to the scheme shown in Table \ref{merging_sequence}.

\begin{table}
\caption{Combinations of nighttime wide fields, rising and setting, observed by AdvACT. Each row shows the two wide fields observed for each night. The numbers indicate the sequence of the observations over a twelve night period. This sequence repeats periodically for the entire season.}\label{merging_sequence}
\begin{center}
\begin{tabular}{ l | c | c | c  }
&\multicolumn{3}{c}{Elevations}\\
\hline
Field combination (two fields per night)  &  $40^{\circ}$   & $45^{\circ}$   & $47^{\circ}$  \\
\hline
wide\tu01h\tu n rising, wide\tu12h\tu n rising   & 1 & 5 & 9         \\
wide\tu01h\tu s rising, wide\tu12h\tu n rising   & 2 & 6 & 10         \\
wide\tu01h\tu n setting, wide\tu12h\tu n setting & 3 & 7 & 11         \\
wide\tu01h\tu s setting, wide\tu12h\tu n setting & 4 & 8 & 12        \\
 \hline
\end{tabular}
\end{center}
\end{table}

\begin{table}
\caption{Summary of areas (sq. deg.) observed by AdvACT in 2016. Note that day\tu 02h\tu s and day\tu 14h\tu overlap with the wide fields.}\label{area_summary_advact}
\begin{center}
\begin{tabular}{ l | c | c | c | c | c}
                     &   wide\tu 01h\tu n  & wide\tu 01h\tu s   & wide\tu 12h\tu n & day\tu 02h\tu s & day\tu 14h\tu n \\
\hline
Season 1 (2016)       &  5700 sq. deg.             &   5000 sq. deg.            &    3400 sq. deg.          &  1700 sq. deg.           & 870 sq. deg.      \\
 \hline
\end{tabular}
\end{center}
\end{table}

We use these twelve strategies to produce uniform and cross-linked coverage for the wide fields by alternating between them, one per night for each twelve night period. To reduce the idle time and the gaps in the observing plan we also observe $deep5$ as a lower priority field, when no other field is available at the desired elevations. The choice of the elevations for $deep5$ is constrained by its low priority. We observe $deep5$ at five elevations: $37^{\circ}$, $39^{\circ}$, $43^{\circ}$, $45^{\circ}$, and $50^{\circ}$.

For daytime observations we select two smaller fields located in low foreground regions. We plan to use these daytime fields for measurements of the CMB B-mode polarization. The motivation for using daytime data in these clean fields is that the large angular scales needed for B-modes measurements are less affected by the diurnal distortion of the beam shape. Moreover AdvACT will take advantage of rotating half-wave plates that will modulate the polarization signal to improve the reconstruction of large angular scale modes. Using multiple daytime fields will allow AdvACT to check the consistency of any measured B-mode signal across different regions of the sky, offering an additional check for galactic foreground contaminations. One of these day-time patches (\textit{day\tu 02h\tu s}) encloses the \textit{deep8} and \textit{deep9} fields already observed during season 3 with ACTPol. It is centered on (RA, Dec) = $(38^{\circ},-40.5^{\circ})$ and covers an area of about $1700$ sq. deg. Since this field is in the cleanest part of the sky and it has already been partially observed we give to it the  highest priority during daytime observations. In Figure \ref{advact_fields} it can be seen that this field is far from the Sun for the entire year. The second field (\textit{day\tu 14h\tu n}) is centered on (RA, Dec) = $(210^{\circ},11^{\circ})$ and covers about $870$ sq.deg. Both daytime fields are observed at elevations of $40^{\circ}$ and $45^{\circ}$. 

This scheme is beneficial for multiple reasons. As indicated above, we will be able to conduct polarization analysis in different relatively low foreground parts of the sky to check the consistency of the signal across the sky. The northern field overlaps with the ACTPol \textit{BOSS-N} field and the \textit{wide\tu 12h\tu n} where we will also have nighttime data. This overlap will allow comparisons between daytime and nighttime data to confirm that daytime beam distortions are handled adequately in the analysis. Finally, these two daytime fields are never directly crossed by the Sun, which will only marginally affect the edges of \textit{day\tu 14h\tu n}  in the last part of the season and only through sidelobes.

Both the daytime and nighttime strategies are optimized to minimize the idle time of the telescope to less than $\sim1.3\%$ of the entire available time. This residual idle time is due to short gaps lasting a few minutes which arise between observations of different fields.

\begin{figure}[!h]
\begin{center}
 \begin{tabular}{c}
\includegraphics[height=8cm]{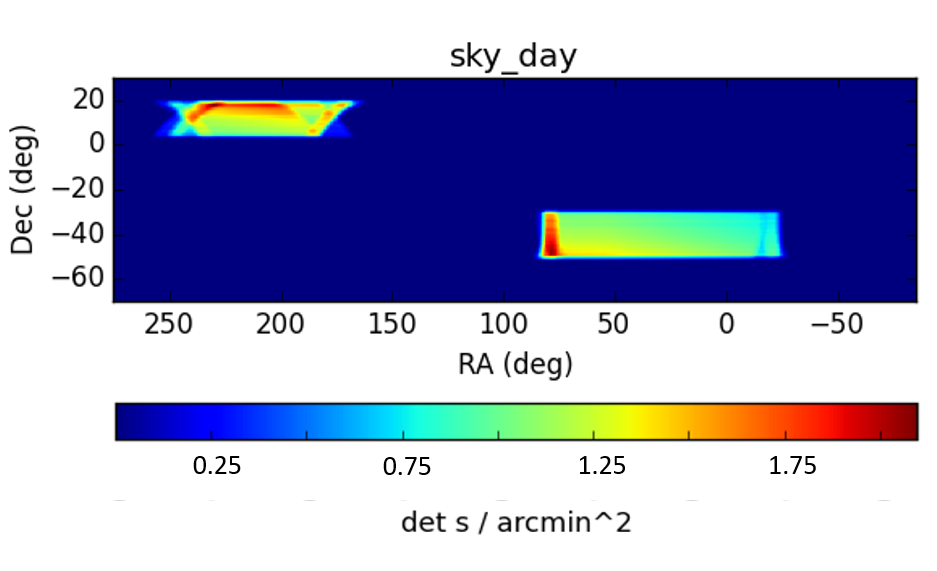}\\
\includegraphics[height=7.5cm]{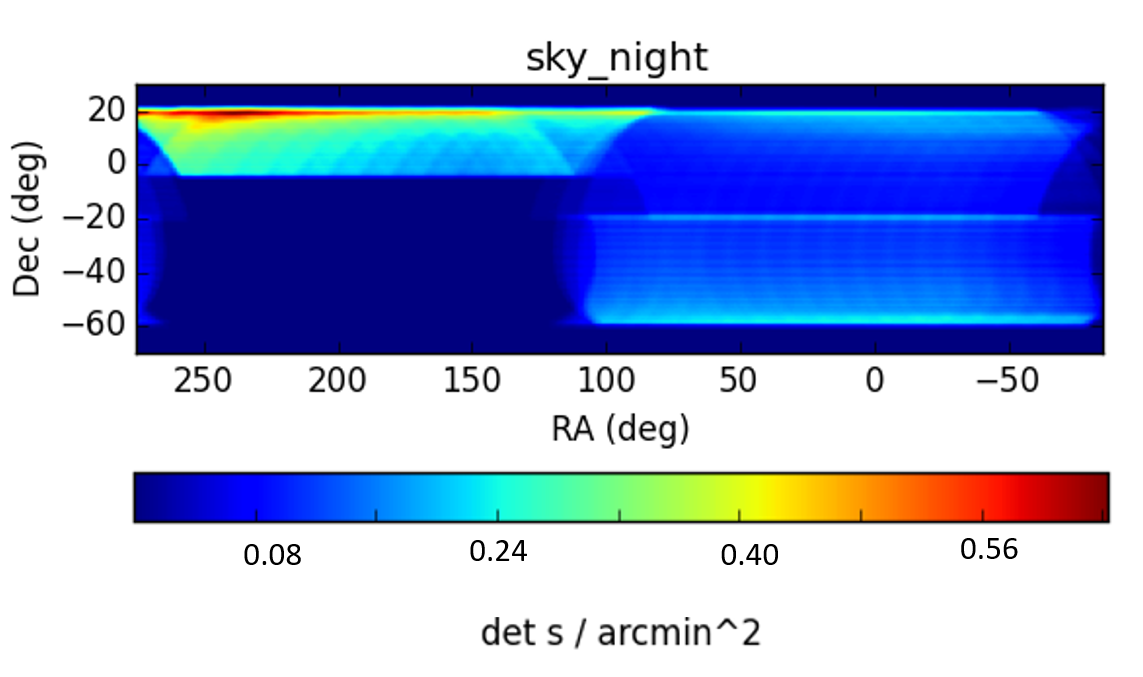}
\end{tabular}
\end{center}
\caption
       {Expected coverage of the AdvACT fields for April-December 2016. \textit{Top}: daytime fields coverage. \textit{Bottom}: nighttime wide fields. It can be seen that all the fields will be cross-linked. During the nighttime we will also observe deep5 (not showed in the plot) with lower priority and because it is visible during the gaps in the primary observing plan. This field has been already observed in the ACTPol seasons. The color scale shows the telescope time per area, i.e. assuming $N=1$ detectors.}\label{depth}
\end{figure}

\begin{figure}[!h ]
\begin{center}
\begin{tabular}{c}
\includegraphics[height=6cm]{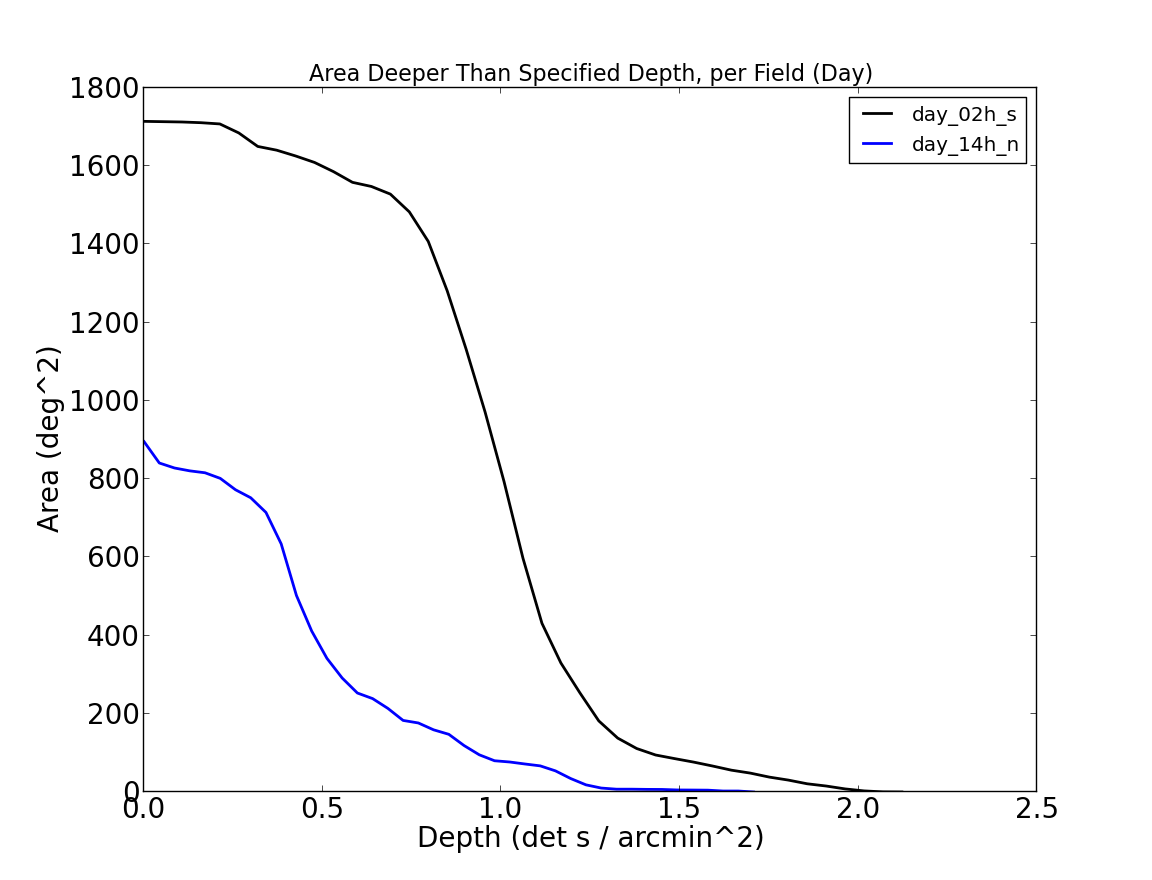}
\includegraphics[height=6cm]{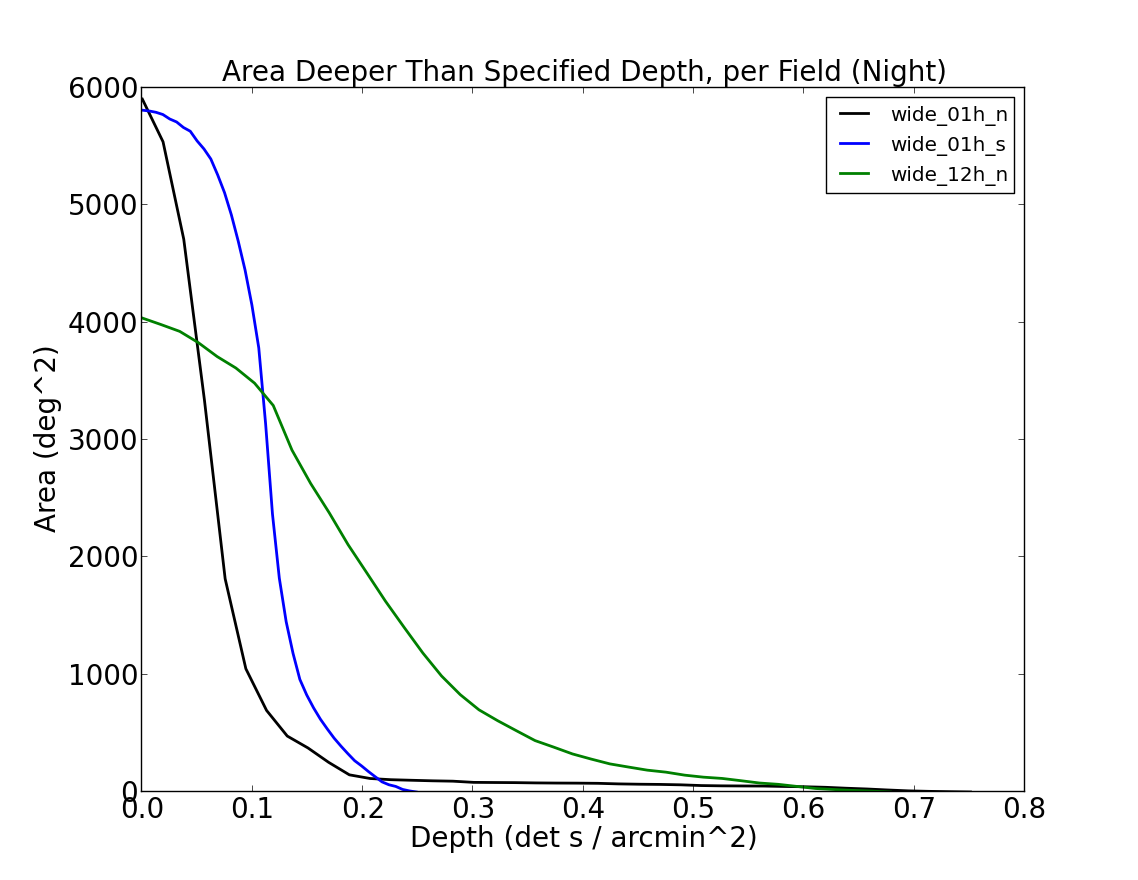}
\end{tabular}
\end{center}
\caption{Area deeper than a given depth for the nighttime and daytime fields for the first AdvACT season.}\label{area_depth}
\end{figure}

In Figure \ref{depth} we show the projected depth maps for the daytime and nighttime AdvACT fields, assuming continuous observations between April and December 2016. In Figure \ref{area_depth} we show the area deeper than a given depth for the wide night fields and the day fields and that \textit{wide\tu 01h\tu s} will be on average deeper than  \textit{wide\tu 01h\tu n} and the same for \textit{day\tu 02h\tu s} compared with \textit{day\tu 14h\tu n}. 

The first season of observations with AdvACT is 2016. The AdvACT survey will observe the CMB sky for at least three years, achieving new levels of angular resolution, sensitivity and frequency coverage, and will overlap with optical surveys, enabling powerful probes of the origin and evolution of the universe.

The optimization of the survey strategy for ACTPol and AdvACT described in this paper can help plan future ground-based CMB surveys, such as for the upcoming Simons Observatory \cite{simons}, also located in the Atacama Desert, and in the longer term, for CMB Stage IV.

\section*{Acknowledgments}
This work was supported by the U.S. National Science Foundation through awards 0408698 and 1440226. The NIST authors would like to acknowledge the support of the NIST Quantum Initiative. The development of multichroic detectors and lenses was supported by NASA grants NNX13AE56G and NNX14AB58G. The work of KPC, KTC, EG, BJK, CM, BLS, JTW, and SMS was supported by NASA Space Technology Research Fellowship awards. MDN acknowledges support from NSF award AST-1454881. DNS acknowledges support from NSF award AST-1311756.

\clearpage

\bibliography{spie_bib}   % bibliography data in report.bib
\bibliographystyle{spiejour}   % makes bibtex use spiejour.bst

\end{document}